\documentclass{ifacconf}

\usepackage{graphicx}      
\usepackage{natbib}        

\usepackage{amsmath,amssymb}
\usepackage{subfigure}
\usepackage{comment,color}
\usepackage{setspace}
\usepackage{booktabs}
\usepackage{subfigmat}

\newcommand{\R}{\ensuremath{{\mathbb R}}}
\newcommand{\Z}{\ensuremath{{\mathbb Z}}}
\newcommand{\N}{\ensuremath{{\mathbb N}}}

\newcommand{\ra}{\rightarrow}

\newcommand{\uuu}{\mathsf{u}}

\newcommand{\nnn}{\mathsf{n}}
\newcommand{\mmm}{\mathsf{m}}
\newcommand{\ppp}{\mathsf{p}}

\newcommand{\sk}{\mathsf{sk}}

\newcommand{\rrr}{\mathsf{r}}

\newcommand{\LLL}{\mathsf{L}}

\newcommand{\uu}{\mathbf{u}}
\newcommand{\yy}{\mathbf{y}}

\newcommand{\ol}{\overline}

\newcommand{\Enc}{\mathsf{Enc}}
\newcommand{\Dec}{\mathsf{Dec}}

\usepackage{xcolor}

\begin{document}
\begin{frontmatter}

\title{
	Toward Nonlinear Dynamic Control over Encrypted Data for Infinite Time Horizon\thanksref{footnoteinfo}}

\thanks[footnoteinfo]{This work was supported in part by
	the National Research Foundation of Korea (NRF) grant funded by the Korea government (Ministry of Science and ICT) (No.~NRF-2017R1E1A1A03070342)
	and in part by the NATO Science for Peace and Security (SPS) PROGRAMME under the project SPS.SFP G5479.
	}

\author[First]{Junsoo Kim} 
\author[Second]{Farhad Farokhi} 
\author[Second]{Iman Shames}
\author[First]{Hyungbo Shim}

\address[First]{ASRI, Dep. of Electrical and Computer Engineering, Seoul National University, Seoul, Korea.}
\address[Second]{Dep. of Electrical and Electronic Engineering, The University of Melbourne, Parkville, VIC 3010, Australia.}

\begin{abstract}
{Recent studies on encrypted control using homomorphic encryption allow secure   operation by directly performing computations on encrypted data without decryption. Implementing dynamic controllers on encrypted data presents unique challenges due to limitations on the number of operations on an encrypted message. Hence, it may not be possible to perform the recursive operations for an infinite time horizon. In this note, we demonstrate that it is possible to run a dynamic controller over encrypted data for an infinite time horizon if the output of the controller can be represented as a function of a fixed number of previous inputs and outputs. The presented implementation requires encryption at both input and output of the plant. We identify a class of nonlinear systems that can accommodate the proposed implementation. The closed-loop performance can be guaranteed using the proposed encrypted controller by ensuring that quantization error is made arbitrarily small with appropriate choice of parameters. We show that the proposed method is amenable to linear systems (as a subset of the said nonlinear systems) with performance guarantees.
}
\end{abstract}


\end{frontmatter}


\section{Encrypted Control}
The notion of encrypted control,
introduced by \cite{Kogiso15CDC}, \cite{Farokhi17CEP}, and \cite{Kim16NECSYS},
aims for security enhancement by performing all the operations in networked controllers directly over encrypted data without decryption.

Let a simple form of static controller be given as
\begin{equation*}
u(t) = h(y(t)),\quad t=0,1,2,\cdots,
\end{equation*}
where $y(t)\in\R^\ppp$ is the input of the controller (the output of the plant),
and $u(t)\in\R$ is the output of the controller (the input of the plant). And, suppose that it is implemented in a digital computer as
\begin{subequations}\label{eq:static_quantized}
\begin{align}
\ol u(t) &= \ol h\left(\left\lceil\frac{y(t)}{\rrr}\right\rfloor\right)=: \ol h(\ol y(t))\in\Z_N,\label{eq:static_integer}\\
u_q(t) &= \LLL \cdot \ol u(t),
\end{align}
\end{subequations}
in which
$\Z_N:=\{i\in\Z: -N/2\le i < N/2\}$ denotes the set of integers from $-N/2$ to $N/2$ with $N\in\N$ chosen sufficiently large,
$\lceil\cdot\rfloor$ is the (component-wise) rounding operation, $\rrr>0$ is the step size for quantization of $y(t)\in\R^\ppp$
so that
the components of $\ol y(t)=\lceil y(t)/ \rrr\rfloor$ are all integers,
$\ol h(\cdot)$
is a function consisting of additions or multiplications so that it results in $\ol u(t)\in\Z_N$,
and
$\LLL>0$ is a scale factor so that the real-valued output $u_q(t)\in\R$, which is approximate to $u(t)$, is obtained from $\ol u(t)$.

The realization of encrypted control is possible when the employed cryptosystem allows arithmetic operations over encrypted data.
For a vector of integers in $\Z_N$,
let $\Enc(\,\cdot\,,\,\sk)$ be the encryption algorithm with secret key $\sk$ (or a public key) so that $\Enc(m,\,\sk)$ is an encrypted message of a vector $m\in\Z_N^\ppp$, $\ppp\ge 1$,
and $\Dec(\,\cdot\,,\sk)$ be the decryption algorithm satisfying
$\Dec(\Enc(m,\,\sk),\,\sk)=m$, $\forall m\in\Z_N^\ppp$.
Suppose that the encryption $\Enc$ is {\it homomorphic} with respect to the function $\ol h$,
i.e.,
there exists an operation ${\bf h(\cdot)}$ over encrypted data such that it satisfies
\begin{equation*}
\Dec({\bf h} (\Enc(\ol y(t),\,\sk)),\,\sk) = \ol h(\ol y(t))
\end{equation*}
for each message $\ol y(t)$ of integers.
Then,
the operation part \eqref{eq:static_integer} can be implemented over encrypted data as
\begin{equation}\label{eq:enc_static}
\uu(t) = {\bf h}(\yy(t)),
\end{equation}
where $\yy(t):=\Enc(\ol y(t),\,\sk)$.
Receiving the encrypted signal $\yy(t)$
and yielding the output $\uu(t)$
from which the real output $u_q(t)$ can be obtained by decryption as $u_q(t) = \LLL \cdot \Dec(\uu(t),\,\sk)$,
it is clear that
the performance of the encrypted controller \eqref{eq:enc_static} is the same with \eqref{eq:static_integer},
while it does not utilize the decryption or the secret key for the control operation part.


The elimination of secret key from the controller is significant in terms of security, so the concept of encrypted control has been considered in various applications, such as model predictive control by \cite{Darup18CSL} or quadratic optimization by \cite{Shoukry16CDC}.

In the meantime, {\it dynamic control} over encrypted data has been a challenge.
Consider a dynamic controller written as
\begin{align}
\begin{split}\label{eq:dynamic}
x(t+1) &= f(x(t),y(t)),\\
u(t) &= h(x(t)),
\end{split}\qquad t=0,1,2,\cdots,
\end{align}
which additionally considers recursive update of the state $x(t)\in\R^\nnn$.
From the rationale that the controller \eqref{eq:dynamic} is supposed to stabilize the closed loop system with the plant, we assume that the norms of the signals $y(t)$, $x(t)$, and $u(t)$ are bounded by a known constant.
One direct way
for implementing \eqref{eq:dynamic} over encrypted data
would be to design functions ${\bf f}$ and $\bf h$ over encrypted data so that they are equivalent to $f$ and $h$, and have the encrypted state recursively updated for each time step, by the function $\bf f$.

However, the recursive operation over encrypted data may not be possible for infinite time horizon
because the number of operations that can be performed on ciphertexts may be limited.
The related issues are as follows:
\begin{itemize}
	\item The employed cryptosystem may not be ``fully homomorphic,'' i.e., it may not be possible to implement all kinds of operations over encrypted data.
	For example, making the use of cryptosystems allowing additions only,
	it is known that the number of multiplications over encrypted data may be limited, especially when it considers multiplications with non-integer rational numbers. See \citep{Cheon18} for more details.
	
	
	\item Even if the employed cryptosystem is fully homomorphic so that all kinds of arithmetic operations can be performed infinite number of times in theory,
	it may not be feasible in practice, due to computational complexity.
	For example, the use of ``bootstrapping'' technique, developed by \cite{Gentry09}, can be considered for the sake of both addition and multiplication unlimited number of times,
	but it may not be easily applied for control systems in practice, if the bootstrapping takes longer than the sampling period.
\end{itemize}

As a consequence,
in terms of encrypted dynamic systems,
most results in literature end up considering finite time dynamic operation~\citep{Alexandru18,murguia2018secure} or end up being satisfied with static operations over encrypted data~\citep{Darup18CSL}.

\section{Toward Unlimited Dynamic Control}

In this short note,
we address that
implementing the operation of
\eqref{eq:dynamic} over encrypted data for infinite time horizon is rendered possible,
if there exist an integer $\mmm\ge 1$ and a function $g$
so that
the system \eqref{eq:dynamic} can be represented as the form
\begin{equation}\label{eq:represented}
\uuu(t) = g(\uuu(t-1),\cdots,\uuu(t-\mmm),y(t-1),\cdots,y(t-\mmm)),
\end{equation}
with the initial values
$\{\uuu(-i)\}^{\mmm}_{i=1}\subset\R$
and $\{y(-i)\}_{i=1}^{\mmm}\subset\R^\ppp$,
so that it satisfies
$\uuu(t)=u(t)$ for all $t\ge 0$.
To see the detail, let us consider encrypting \eqref{eq:represented};
suppose that it is implemented over integers as
\begin{align}
\ol \uuu'(t) &= \ol g(\ol\uuu(t-1),\cdots,\ol\uuu(t-\mmm),\ol y(t-1),\cdots,\ol y(t-\mmm)),
\notag\\
\uuu_q(t) &= \LLL\cdot \ol \uuu'(t),\qquad
\ol \uuu(t) = \left\lceil\frac{\uuu_q(t)}{\rrr}\right\rfloor\label{eq:represented_quantized}
\end{align}
where $\ol \uuu(-i)=\lceil\uuu(-i)/\rrr\rfloor$ and $\ol y(-i)=\lceil y(-i)/\rrr\rfloor$, $i=1,\cdots,\mmm$, are the initial values,
$\ol y(t)= \lceil y(t)/\rrr\rfloor$, $t\ge 0$, is the quantized input,
the function $\ol g$ consisting of integer addition and multiplication yields the output $\ol \uuu'(t)\in\Z_N$ so that the approximate output $\uuu_q(t)\in\R$ is obtained by scaling with the factor $\LLL>0$,
and the output $\ol \uuu(t)$, as an integer of scale $1/\rrr$, is fed back to the controller at each time $t$.

Then,
we suggest that the operation of \eqref{eq:represented_quantized} can be implemented over encrypted data to operate for infinite time horizon.
In encrypted control systems, it is expected that the real-valued output $\uuu_q(t)$ would be obtained at the actuator with decryption,
and
it would be able to re-encrypt and transmit the encryption of $\ol \uuu(t)$ to the controller, as a newly encrypted message.
As a result,
it will be able to implement \eqref{eq:represented_quantized} over encrypted data as
\begin{align}
\uu(t) &= {\bf g}\left(\left\{\Enc\left(\ol \uuu(t-i),\sk\right)
,~
\Enc\left(\ol y(t-i),\sk\right) \right\}_{i=1}^{\mmm}\right)
\label{eq:represented_encrypted}
\end{align}
in which the encrypted controller receives both the signals $\ol \uuu(t)$ and $\ol y(t)$ as newly encrypted messages,
and computes the function $\bf g$ over encrypted data which corresponds to the function $\ol g$ over integers.
Then, the operation of \eqref{eq:represented_encrypted} will not be recursive
with respect to the number of operations for the encrypted messages, since the result $\uu(t)$ will not be re-used for the operation, but to be transmitted to the actuator and decrypted.
Hence, it is able to operate for infinite time horizon as long as the cryptosystem is homomorphic with respect to the function $\ol g$.

Finally,
the following proposition suggests that the encrypted controller \eqref{eq:represented_encrypted} can operate for infinite time horizon and its performance is the same with the quantized controller \eqref{eq:represented_quantized}.

\begin{prop}\label{prop:1}
	Suppose that the encryption $\Enc$ is homomorphic with respect to the function $\ol g$.
	Then, the encrypted controller \eqref{eq:represented_encrypted} can operate for infinite time horizon, i.e., it guarantees $\LLL\cdot\Dec(\uu(t),\sk)=\uuu_q(t)$, for all $t\ge 0$.\hfill$\square$
\end{prop}

	\begin{rem}
	Note that, in Proposition~\ref{prop:1},
	the requirement for the employed cryptosystem is to be homomorphic for newly encrypted messages only (with respect to $\ol g$),
	rather than requiring that the encrypted messages can be used for the operation infinitely many times.
	In particular,
	there is no need of the use of bootstrapping techniques of fully homomorphic cryptosystems.\hfill$\square$
	\end{rem}

\section{Further Issues and Examples}
The proposed approach would be applicable when
the given controller \eqref{eq:dynamic} can be converted to the form \eqref{eq:represented}.
In addition, we list several further issues that should be considered:
\begin{itemize}
	\item
	For the conversion of \eqref{eq:dynamic} to the form \eqref{eq:represented} and \eqref{eq:represented_quantized},
	the function $g$ should be found so that the performance of \eqref{eq:represented} should be equivalent to that of \eqref{eq:dynamic}.
	And,
	in order to have the function $\ol g$ consist of additions and multiplications over integers,
	polynomial approximation methods should be considered for the function $g$, in practice.

	\item
	The effect of quantization errors.
	Considering that the closed loop of the plant and the given controller \eqref{eq:dynamic} is supposed to be stable with respect to external disturbances, such as the errors caused by quantization of signals,
	the converted controller \eqref{eq:represented} should satisfy the same stability so that it can be implemented as \eqref{eq:represented_quantized} over quantized signals, in practice.
\end{itemize}

In the rest of this note,
we suggest a class of nonlinear systems
under which the above issues can be resolved.
First, the following assumption asks that the given controller can be transformed into ``observable canonical form''.
\begin{assum}\label{asm:obs}
There exist  $\nnn'\le \nnn$ and a diffeomorphism $\Phi:\R^\nnn\ra\R^{\nnn'}\times\R^{\nnn-\nnn'}$ such that,
by $[z^\top,z'^\top]^\top:=\Phi(x)$,
the system \eqref{eq:dynamic} is transformed into the form
\begin{align}
z(t+1) &= \begin{bmatrix}
z_1(t+1)\\z_2(t+1)\\z_3(t+1)\\\vdots\\z_{\nnn'}(t+1)
\end{bmatrix}
=
\begin{bmatrix}
g_1(z_{\nnn'}(t),y(t))\\g_2(z_1(t),z_{\nnn'}(t),y(t))\\g_3(z_1(t),z_2(t),z_{\nnn'}(t),y(t))\\\vdots
\\ g_{\nnn'}(z(t),y(t))
\end{bmatrix}\notag\\
z'(t+1)& = g'(z(t),z'(t),y(t))\label{eq:observable}\\
u(t) &= z_{\nnn'}(t).\notag
\end{align}
\hfill$\square$
\end{assum}

Then, the following proposition states that the system \eqref{eq:observable} can be implemented as the form \eqref{eq:represented}, with the same performance.

%

\begin{prop}
Under Assumption~\ref{asm:obs},
there exists a function $g$ with $\mmm=\nnn'$, such that the operations of \eqref{eq:represented} and \eqref{eq:observable} are identical, i.e., it satisfies $\uuu(t) = u(t)$, $\forall t\ge 0$.
\hfill$\square$
\end{prop}

Moreover, the following corollary suggests that the system \eqref{eq:observable} can be implemented over integers and quantized signals as in \eqref{eq:represented_quantized},
when the functions $\{g_i\}_{i=1}^{\nnn'}$ are of real polynomials (or able to be approximated with real polynomials),
where the performance error due to quantization can be made arbitrarily small with the choice of parameters.
And then, according to Proposition~\ref{prop:1},
it will be able to be implemented over encrypted data with the same performance as well, and able to operate for infinite time horizon without decryption.
\begin{cor}\label{cor:performance}
Suppose that Assumption~\ref{asm:obs} holds, and the functions $\{g_i\}_{i=1}^{\nnn'}$ are of real polynomials.
Then, for every $\epsilon>0$, there exist $\rrr'>0$, $\LLL'>0$, and a polynomial $\ol g$
of integer coefficients
such that for every $\rrr<\rrr'$ and $\LLL<\LLL'$, the quantized controller \eqref{eq:represented_quantized}
guarantees $\|\uuu_q(t)- u(t)\|<\epsilon$, for all $t\ge 0$.
Furthermore, the encrypted controller \eqref{eq:represented_encrypted} guarantees $$\|\LLL\cdot\Dec(\uu(t),\sk)- u(t)\|<\epsilon,$$ for all $t\ge 0$.
\hfill$\square$
\end{cor}

Finally,
the following proposition suggests that
every linear system satisfies the condition of Assumption~\ref{asm:obs},
so that the proposed method is applicable for all linear systems.
\begin{prop}\label{prop:linear}
	Given the system \eqref{eq:dynamic},
	suppose that $f$ and $h$ are of linear functions.
	Then, there exists an invertible matrix $T\in\R^{\nnn\times\nnn}$ such that, by $[z^\top,z'^\top]^\top=Tx$, the system \eqref{eq:dynamic} is transformed into the form \eqref{eq:observable},
	where $\{g_i\}_{i=1}^{\nnn'}$ and $g'$ are of linear functions.\hfill$\square$
\end{prop}
\begin{rem}
From the combination of Corollary~\ref{cor:performance} and Proposition~\ref{prop:linear}, it is clear that all linear systems can be encrypted to operate for infinite time horizon, in which the performance error due to quantization can be made arbitrary small.
It accords with the previous result in \cite{KimTAC20},
but it should be noted that the results of this paper based on Proposition~\ref{prop:1} do not require that the encryption algorithm should be homomorphic for infinitely many times of operations (such as recursive operations). Instead, it requires the cryptosystem to be homomorphic only for finitely many additions or multiplications.\hfill$\square$
\end{rem}
	
	\section{Conclusion}
	In this short note, we have reviewed the challenge of implementing dynamic controllers over encrypted data in terms of operating for infinite time horizon.
	Considering that the output of the encrypted controller is supposed to be decrypted at the actuator so that it would be able to be re-encrypted and transmitted the the controller again,
	we suggest that
	dynamic control operations over encrypted data
	would be possible for infinite time horizon when the output of the system can be represented as a function of the previous inputs and outputs.
	Then, it is suggested that systems that are able to be represented as nonlinear observable canonical form can be implemented over encrypted data to operate for infinite time horizon, in which the performance error can be made arbitrarily small with the choice of parameters.

\bibliography{ifacconf}             
                                                   








\end{document}